# A Deep Reinforcement Learning Approach to Automated Stock Trading, using xLSTM Networks

Faezeh Sarlakifar[1] Mohammadreza Mohammadzadeh Asl[1] Sajjad Rezvani Khaledi[2] Armin Salimi-badr[1]

[1] *Faculty of Computer Science and Engineering, Shahid Beheshti University, Tehran, Iran*
[2] *Faculty of Electrical Engineering, Shahid Beheshti University, Tehran, Iran*

CORRESPONDING AUTHOR: ARMIN SALIMI-BADR (e-mail: a_salimibadr@sbu.ac.ir)

***Abstract*—Traditional Long Short-Term Memory (LSTM) networks are effective for handling sequential data but have limitations such as gradient vanishing and difficulty in capturing long-term dependencies, which can impact their performance in dynamic and risky environments like stock trading. To address these limitations, this study explores the usage of the newly introduced Extended Long Short-Term Memory (xLSTM) network in combination with a deep reinforcement learning (DRL) approach for automated stock trading. Our proposed method utilizes xLSTM networks in both actor and critic components, enabling effective handling of time series data and dynamic market environment. Proximal Policy Optimization (PPO), with its ability to balance exploration and exploitation, is employed to optimize the trading strategy. Experiments were conducted using financial data from major tech companies over a comprehensive timeline, demonstrating that the xLSTM-based model outperforms LSTM-based methods in key trading evaluation metrics, including cumulative return, average profitability per trade, max earning rate, maximum pullback, and Sharpe ratio. These findings mark the potential of xLSTM for enhancing DRL-based stock trading systems.**

***Keywords*—***Extended Long Short-Term Memory (xLSTM), Proximal Policy Optimization (PPO), Automated Stock Trading, Actor-Critic Reinforcement Learning*

## I. INTRODUCTION

In general, stakeholders aim to maximize their returns by predicting market trends. However, there is always concern about achieving maximum profit in a highly complex and fluctuating market environment. This is challenging because human insights are limited and cannot easily account for all relevant factors. To address this, researchers have focused on developing automated trading systems that can adapt to market changes and make predictions from a more generalized perspective [1].

Previously, many research studies on the development of automated stock trading systems employed supervised learning approaches [2]. Although these approaches were impressive, they had many limitations in achieving the optimal point. Over the years, it has become evident that the deep reinforcement learning (DRL) approach has demonstrated promising results in predicting stock prices. Compared to merely using supervised methods like recurrent neural networks (RNN) that lack a mechanism to take actions based on prediction, DRL approaches can dynamically adjust actions in response to new states in the market environment. Moreover, regarding the exploration mechanism in DRL, the model receives feedback from the market, and by employing it, the model becomes more capable of exploiting profitable patterns.

While some early works use deep Q-learning models [3], among reinforcement learning (RL) algorithms, deep Q-learning (DQL) is often considered less robust compared to more advanced approaches, such as Proximal Policy Optimization (PPO), which addresses key challenges in stability and sample efficiency.

In other research studies, an agent-driven model has been proposed, enhanced by DRL and imitation learning, which enables the agent to learn directly from raw inputs and perform better in high-complexity problems. In this study, the trading process is formulated as a partially observable Markov decision process (POMDP) to account for the noisy nature of financial data [4].

There is an inspiring research study, focused on using cascaded LSTM networks with a deep reinforcement learning approach. This study combines proximal policy optimization (PPO) with LSTM networks to create a robust and reliable automated stock trading system. This work uses cascaded LSTM networks with DRL where the first LSTM extracts time-series features from daily stock data, and two other LSTM networks are used within the DRL agent for strategy learning [5].

A recent work combined Deep Q-Network (DQN) and Deep Deterministic Policy Gradient (DDPG) with Convolutional Neural Network (CNN) and Gated Recurrent Unit (GRU) architectures for automated stock trading. Additionally, the attention mechanism was added to handle the limitations of RNN networks–GRU [6].

However, with the introduction of extended LSTM (xLSTM) architecture [7], evaluating its performance in automated stock trading with DRL needs to be explored.

xLSTM has mitigated some weaknesses in LSTM models such as gradient vanishing, resulting in forgetting long-term patterns. xLSTM also demonstrates better results than the Transformer architecture in some benchmarks. In a more general context, the application of xLSTM in deep reinforcement learning approaches has been minimally explored and is limited to a single research study [8].

In this study, we utilized the newly introduced xLSTM architecture in combination with DRL to predict stock market prices. xLSTM networks have been used in the RL-based model to retrieve the history of observations in both actor and critic parts, enabling the model to use the retrieved observations to revise its strategy or make predictions.

### A. *Preliminary*

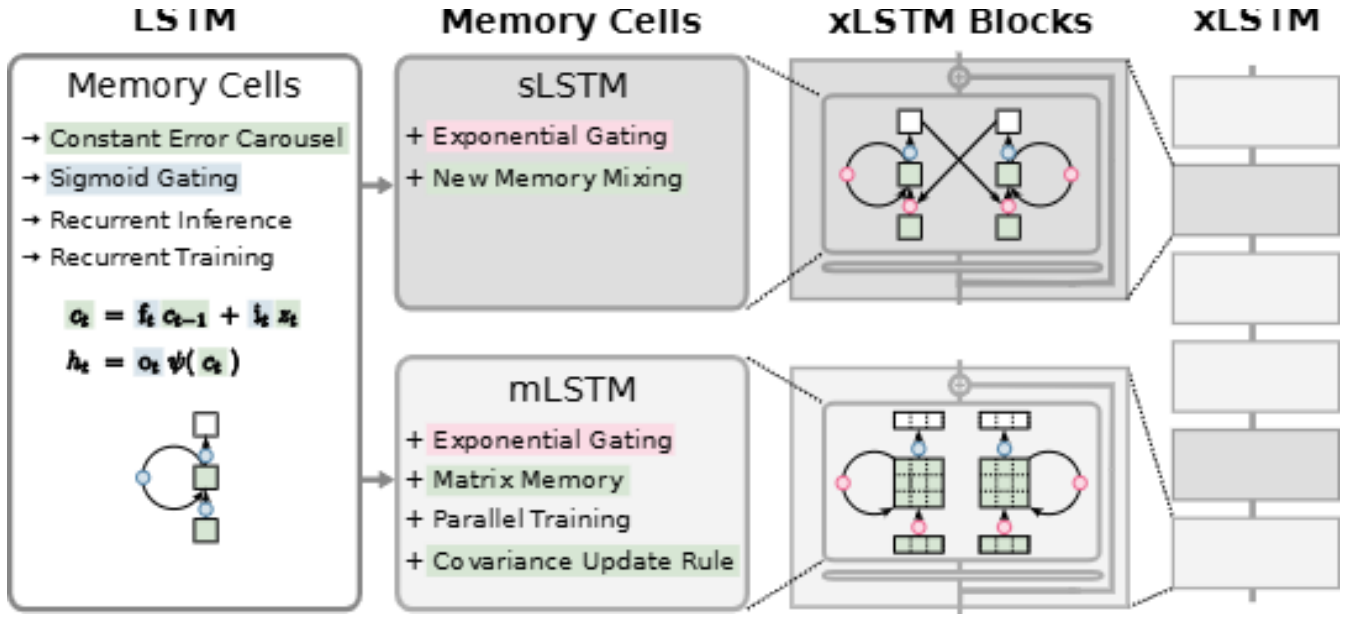

Fig. 1. The extended LSTM (xLSTM) family Architecture [7].

LSTM has been a breakthrough in natural language processing tasks as a result of its capabilities to retrieve the information it has received. However, with more advancements in natural language processing over time, other methods like Transformer architecture were introduced and dealt with some of the key challenges of Recurrent Neural Networks (RNNs). For example, RNNs perform well with sequential and time-series data, but they would either face gradient vanishing or exploding problems for long sequences. Transformers solved this problem by suggesting inherent parallelism and learning both short-term and long-term dependencies, but the cost of this approach was the high parameter usage.

Roughly 27 years after the publication of the original LSTM paper, xLSTM has been introduced to refine LSTM's weaknesses. xLSTM has two major modifications compared to LSTM which are the utilization of exponential gating and the new memory structure. xLSTM architecture is composed of sLSTM and mLSTM blocks as shown in Fig. 1. sLSTM presents scaler memory and update and a novel memory mixing, while mLSTM is designed to be completely parallelizable using matrix memory and covariance update rule. A combination of sLSTM and mLSTM blocks can be selected to create a stack of xLSTM blocks, performing similar tasks to LSTM but with a distinct internal design.

## II. PROPOSED METHOD

In this section, we provide a brief overview of our proposed approach. First, introduce our general architecture and describe the relationships between the main building blocks. We then explain each block, followed by an outline of our reinforcement learning algorithm and the training description of xLSTM networks. Finally, we describe our stock trading environment and the reward function that we employ in our environment.

### A. *General Model Architecture*

Our proposed method is based on the Proximal Policy Optimization (PPO), one of the best-performing reinforcement learning algorithms. We have used Recurrent PPO from Stable Baselines3 library[1] which has added support for recurrent policies to implement PPO. We have implemented a new RecurrentActorCriticPolicy named xLSTMPolicy, connected this module to the Recurrent PPO module, and worked together as a DRL model. The xLSTMPolicy utilizes the official xLSTM library[2].

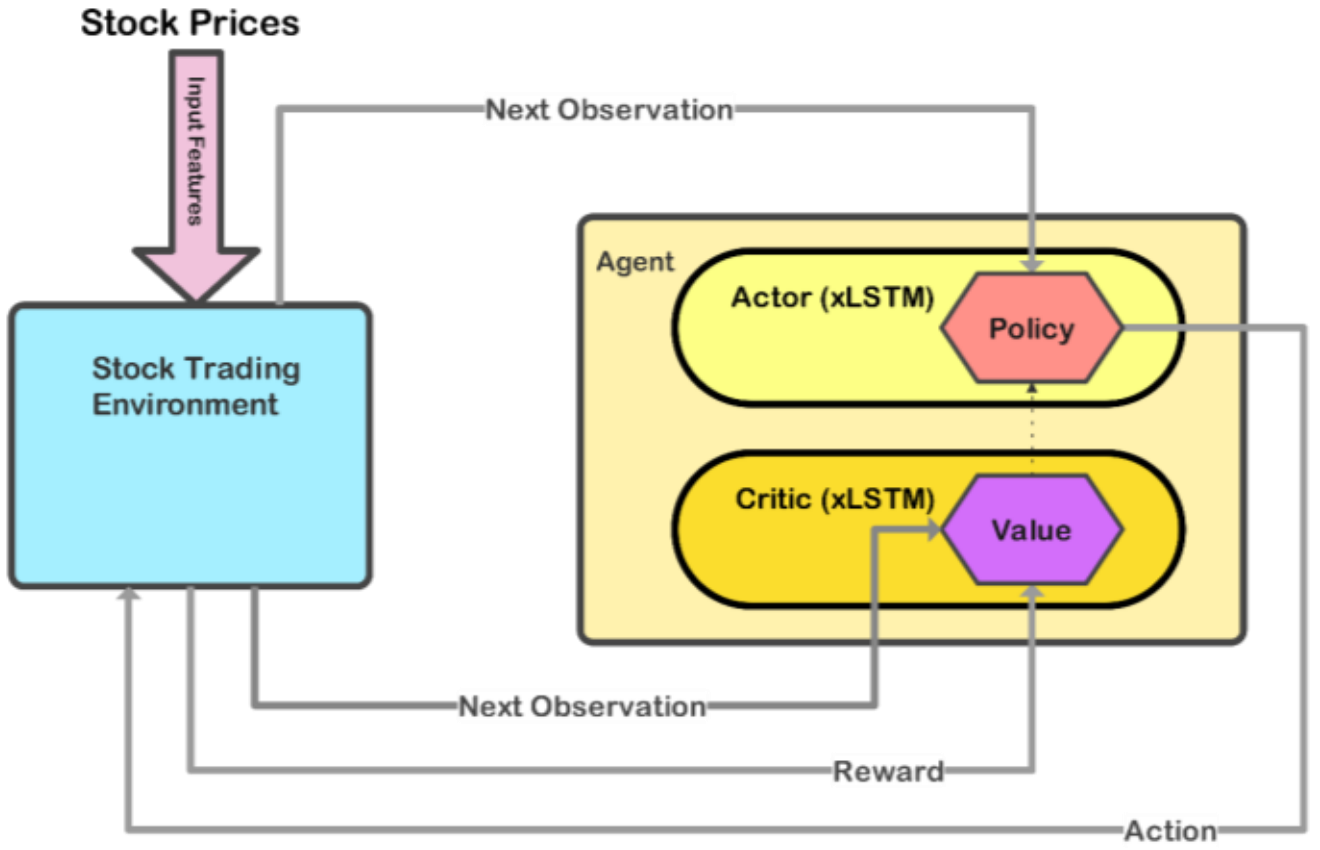

Fig. 2. General Proposed Model Architecture.

### B. *Recurrent Proximal Policy Optimization (PPO)*

Proximal Policy Optimization (PPO) is a popular algorithm in reinforcement learning that balances exploration and exploitation while optimizing a policy. This algorithm is implemented in the stable-baseline3 library with recurrent neural networks (LSTM) to define a powerful and robust pipeline for processing sequential data and making reliable predictions–Recurrent PPO. In this study, we utilized xLSTM networks instead of LSTM to test whether it is effective to use or not, in the context of time series data and stock trading tasks.

Algorithm 1, starts with randomly initializing two xLSTM neural networks–an actor network for making decisions and a critic network for estimating the value of states. It then trains these networks through repeated interactions with our stock trading environment. During each episode, the agent observes financial states, decides on actions, receives rewards, and progressively improves its decision-making strategy by learning from these interactions. The xLSTM architecture allows the agent to maintain memory of past states, which is

[1] https://sb3-contrib.readthedocs.io/en/master/modules/ppo_recurrent.html

[2] Available at: https://github.com/NX-AI/xlstm

crucial for understanding complex financial time series and making sequential investment or trading decisions.

**Algorithm 1**: Recurrent PPO, using xLSTM Actor-Critic Policy

**Inputs:**

- *Initial observation $s_t$*
- *Adam optimizer with learning rate α*
- *Discount factor γ*
- *Clipping range ϵ Advantage estimate $A_t$*
- *Action space*
- *Environment with financial data*
- *xLSTM internal states*

**Outputs:**

- *Parameters θ of xLSTM actor*
- *Parameters φ of xLSTM critic*
- *Trained xLSTM model*

*1* Choose the implemented **xLSTM Recurrent Actor-Critic Policy** as policy_aliases and use its methods and components in this Recurrent PPO module;

*2* Randomly initialize xLSTM actor and critic with parameters θ and φ;

*3* Initialize the replay buffer D;

*4* Initialize xLSTM hidden state $h_0$ and cell state $c_0$;

*5* **for** each episode **do**

*6* Initialize the environment with the initial state $s_0$;

*7* Initialize xLSTM hidden state $h_0$ and cell state $c_0$;

8 **for** each step t in the episode **do**

*9* Receive state $s_t$ from the environment;

*10* Extract features using the MLP feature extractor;

*11* Encode features for xLSTM processing;

*12* Process $s_t$ with xLSTM actor to obtain feature vector $f_t$;

*13* Compute the critic's value estimate $\hat{v}_t = V_{xlst}(f_t)$;

*14* Sample an action $a_t$ from the policy $\pi_{xlst}(a_t \mid f_t)$;

*15* Execute action $a_t$ in the environment to receive the reward $r_t$ and next state $s_{t+1}$;

*16* Process $s_{t+1}$ with xLSTM actor to obtain feature vector $f_{t+1}$;

*17* Compute the advantage estimate $A_t = r_t + \gamma \hat{v}_{t+1} - \hat{v}_t$;

*18* Add the transition $(f_t, a_t, A_t)$ to the replay buffer D;

*19* **if** the episode is terminated **then**

*20* Update the critic: $\varphi \leftarrow \varphi - \alpha V \nabla \phi (r_t + \gamma \hat{v}_{t+1} - \hat{v}_t)^2$;

*21* Update the actor using the PPO objectiv*e function:* $\theta \leftarrow \theta + \alpha\theta \nabla\theta$ L_PPO(θ);

*22* Update hidden state $h_t$ and cell state $c_t$ of xLSTM by backpropagating the gradient of the reward function;

*23* Clear the replay buffer D;

*24* **end if**

*25* **end for**

*26* **end for**

### C. xLSTM Networks

We have utilized two xLSTM networks in our proposed method: one as a policy network and the other as a value network. These networks share the same configuration and architecture, with the Gaussian Error Linear Unit (GeLU) as their activation function during training. The embedding vector size is 128. Our policy and value network work together and interact with the environment to select the most profitable action in each time step as shown in Fig. 2.

### D. Stock Trading Environment

As shown in Fig. 2, our stock trading environment is developed to take the selected action (based on policy) and stock prices from the dataset and return the next observation and reward value to interact with the agent. For more details about this environment, we refer to the initial balance amount, which is $1 million to become compatible with the J. Zou et al. research study [5].

As shown in Algorithm 2, our reward function first checks the market's turbulence index against a predefined threshold to avoid high-risk situations. We penalize the model for choosing actions in unstable market conditions. The penalty value is set as a large, negative value of -1. When the market is stable, our reward function calculates the reward by measuring the total portfolio value change and subtracting transaction costs, then normalizes this raw reward.

**Algorithm 2**: Reward Function

**Inputs**:

- *turbulence_index*
- *turbulence_threshold*
- *current_stock_prices*
- *prev_total_value*
- *balance*
- *shares_held*
- *action*

**Output:** *Normalized_Reward*

*1* Penalty_vlaue = -1;

*2* // Turbulence Check (Market Risk Assessment)

*3* **if** *turbulence_index* > *turbulence_threshold* **then**

*4* *Normalized_Reward* = Penalty_value;

*5* **return** *Normalized_Reward*;

*6* **end if**

*7* *total_value* = *balance* + Sum(

*8* *shares_held*[stock] * *current_stock_prices*[stock]

*9* **for** each stock );

*11* value_change = *total_value - prev_total_value;*

*12* transaction_cost = Compute_Transaction_Cost(*action*, *current_stock_prices*);

*13* raw_reward = value_change - transaction_cost;

*14* *Normalized_Reward* = Normalize_Reward(raw_reward);

*15* **return** *Normalized_Reward;*

## III. EXPERIMENT & RESULT

### A. Dataset

We use market data from Yahoo for five major companies in the tech industry including NVIDIA, Apple, Microsoft, Google, and Amazon. For these companies, we've selected the timeframe from 2009/01/01 to 2022/01/01 as the training data and the timeframe from 2022/01/02 to 2024/04/01 as the test data. For each of these companies, we have low, high, open, close, adj close, and volume prices in our feature set per each day in the chosen range. We also define the turbulence index as (EQ. 1) to avoid trading in extreme market situations [9].

### B. Evaluation Metrics

a. *Cumulative Return (CR)*: This measures the total return of the portfolio after completion of the trading process.

$$CR = \frac{P_{final} - P_{init}}{P_{init}} \quad (1)$$

Where: $P_{final}$ is the final portfolio value, and $P_{init}$ is the portfolio's initial value.

b. *Max Earning Rate (MER)*

This evaluation metric shows the best performance that a trading strategy could achieve over time.

$$MER = max(\frac{A_t - inital_balance}{initial_balance}) \quad (2)$$

Where: $A_t$ is the total asset of strategy at time t.

c. *Maximum PullBack (MPB)*

It measures the maximum percentage of decrease in profitability.

$$MPB = max(\frac{A_{trough} - A_{peak}}{A_{peak}}) \quad (3)$$

Where: $A_{peak}$ is the highest total asset value before a decline, and $A_{trough}$ is the lowest total asset value after the peak, during the decline.

d. *Average Profitability Per Trade (APPT)*

$$APPT = \frac{P_{final} - P_{init}}{n} \quad (4)$$

Where: n is the number of trades.

e. *Sharp Ratio (SR)*

$$SR = \frac{E(R_{portfolio}) - R_{risk-free}}{\sigma_{portfolio}} \quad (5)$$

Where: $E(R_{portfolio})$ is the portfolio's expected return over a given period, $R_{risk-free}$ is the risk-free rate, and $\sigma_{portfolio}$ is the standard deviation of the portfolio's returns, a measure of risk or volatility.

### C. Compared Models

To test the performance of xLSTM in comparison with classic LSTM networks, we trained a base model using Recurrent PPO with MLPPolicy, which employs LSTM networks for both the actor and the critic. The performance of this base model with three different time window sizes–30, 15, and 5–for the test timeframes is shown in Fig. 4.

Another experiment was conducted to find the best hyper-parameters for training our proposed model. In Fig. 3, we can see the return of trained models with different batch sizes, using a time window of 3. This plot shows that using a batch size of 32 generally achieves a higher return. The fluctuations occurred due to the small size of the time window. A long-term view of the dataset and a better examination of historical data can partially solve this problem. Therefore, we trained our models with larger time windows to address this issue.

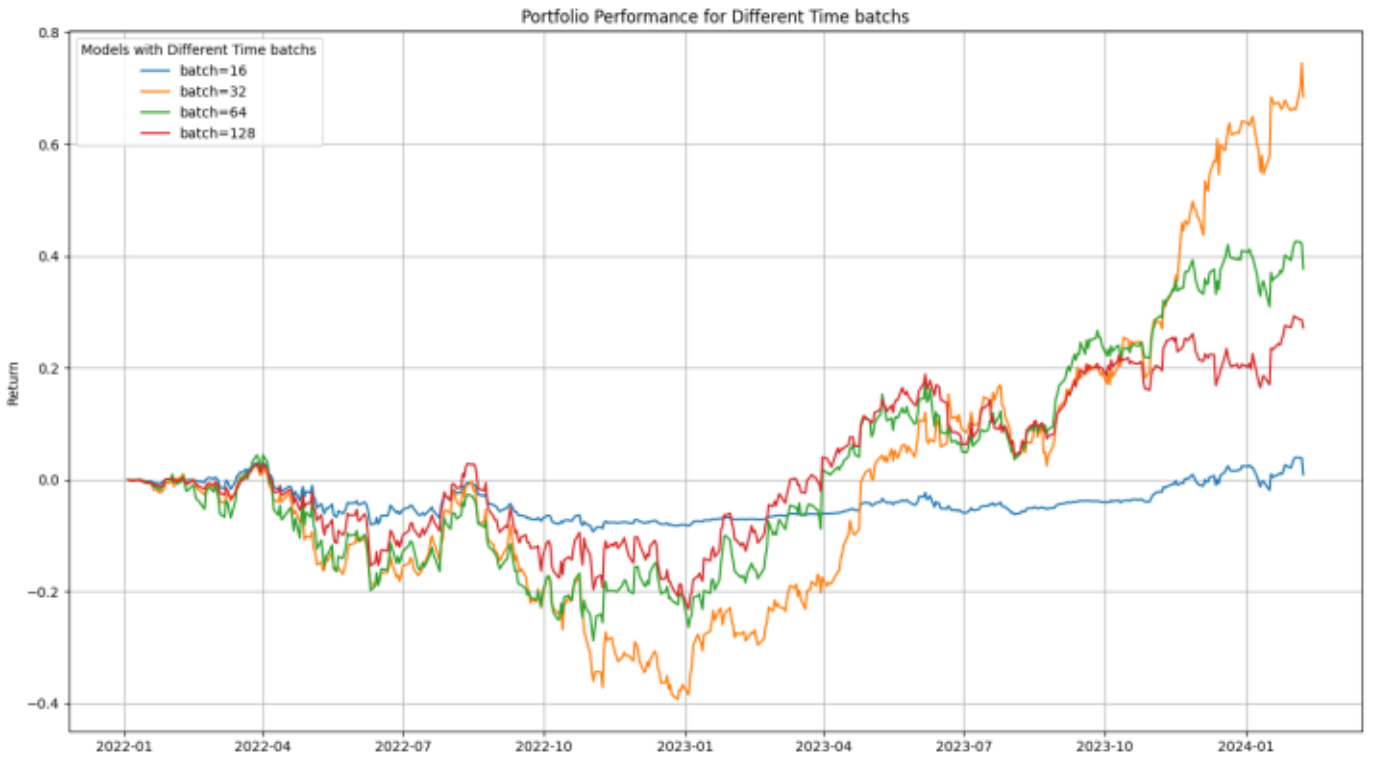

Fig. 3. The trading results of our proposed model with time_window= 3, and different batch sizes.

Fig. 5 presents the return over time for the test data predictions of our proposed model using three different time windows. It is evident that with a time window of 30, the model achieves better outcomes and exhibits a smoother flow. In Fig. 5, negative returns are almost nonexistent, which indicates that this strategy is profitable with minimal risk of financial loss. In the comparison between Fig. 4 and Fig. 5, we can observe the effect of using xLSTM as policy and value networks to achieve more reliable and robust predictions. The smooth flow achieved with xLSTM using a time window size of 30 underscores its potential to inspire further experiments and extend its application in various automated trading strategies with the DRL approach.

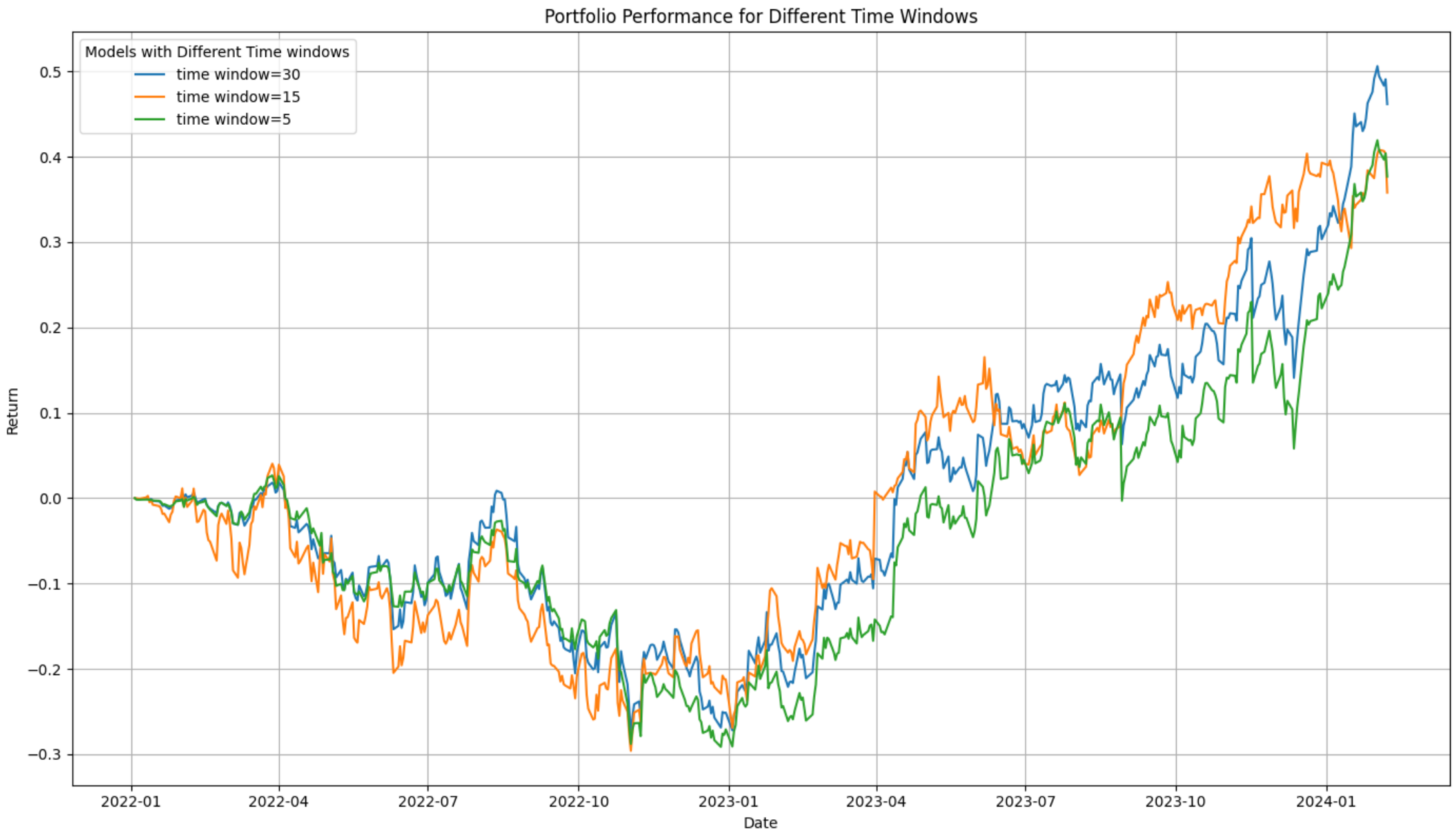

Fig. 4. The trading results of the base model (Recurrent PPO with LSTM networks) with different time windows.

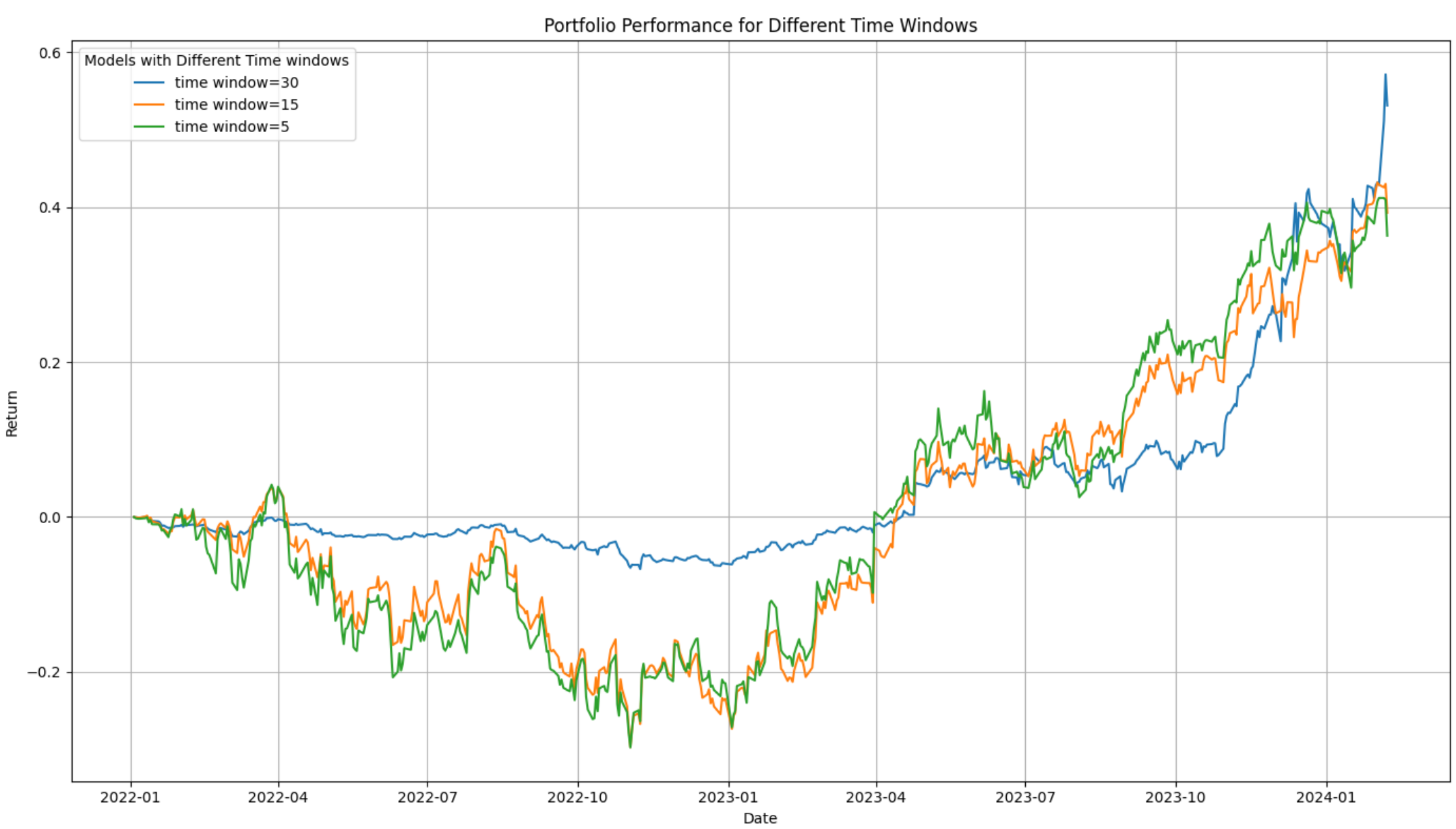

Fig. 5. The trading results of our proposed model with different time windows and batch_size=32.

TABLE I. COMPARISON BETWEEN THE BASE AND OUR PROPOSED MODEL IN IMPORTANT EVALUATION METRICS.

| **Model** | **Configuration** | | **Evaluation Metrics** | | | | |
|---|---|---|---|---|---|---|---|
| | ***Time Window Size*** | ***Batch Size*** | ***CR*** | ***MER*** | ***MPB*** | ***APPT*** | ***SR*** |
| Base (Recurrent PPO with LSTM Networks) | 30 | 64 | 46.16 | 50.61 | 28.92 | 152.58 | 0.799 |
| Base (Recurrent PPO with LSTM Networks) | 15 | 64 | 35.80 | 40.80 | 32.33 | 118.33 | 0.634 |
| Base (Recurrent PPO with LSTM Networks) | 5 | 64 | 37.68 | 41.93 | 30.98 | 124.54 | 0.733 |
| **Proposed Model** | **30** | **32** | **53.11** | **57.12** | **7.58** | **175.54** | **1.650** |

| Model | Configuration | | Evaluation Metrics | | | | |
|---|---|---|---|---|---|---|---|
| | *Time Window Size* | *Batch Size* | *CR* | *MER* | *MPB* | *APPT* | *SR* |
| Proposed Model | 15 | 32 | 39.28 | 43.17 | 32.28 | 129.82 | 0.718 |
| Proposed Model | 5 | 32 | 36.30 | 41.18 | 32.55 | 119.99 | 0.639 |

The evaluation metrics and configurations for both our base and proposed models are presented in Table I. This table compares the performance of using LSTM and xLSTM networks across key trade evaluation metrics. The results clearly demonstrate that incorporating xLSTM into our architecture outperforms the classic LSTM networks across all evaluation metrics.

## IV. Conclusion

This study explored the potential of xLSTM networks combined with the Deep Reinforcement Learning (DRL) approach for automated stock trading.

Our results clearly show that xLSTM outperforms classic LSTM networks, which aligns with the primary goal of this research. xLSTM networks effectively address many LSTM limitations. However, training the xLSTM network requires more computational resources, which makes it challenging to test it on a large-scale problem. Therefore, we begin by evaluating the performance of this key idea using five stock market prices and lightweight features in combination with a popular RL algorithm (PPO) to achieve our goal and explore the capabilities of xLSTM networks for processing time series data.

### *A. Feature Work*

While our results demonstrate the capability of xLSTM in this context, there are some directions for future work to enhance and expand the exploration of this approach.

One key idea for our future work is more powerful feature engineering because meaningful features can effectively help the model perform better and design a better trading strategy. Another idea is to use ensemble modeling for xLSTM networks even in the actor or critic network.